\documentclass[aps,prl,reprint,notitlepage,twocolumn,superscriptaddress]{revtex4-1}

\usepackage{amsmath,amssymb}
\usepackage{mathtools}
\usepackage{float}
\usepackage{graphicx}
\usepackage{dcolumn}
\usepackage{bm}
\usepackage{caption}
\usepackage{subcaption}
\usepackage{etoolbox}
\usepackage{tikz}
\newrobustcmd*{\mycircle}[1]{\tikz{\filldraw[draw=#1,fill=#1] (0,0) circle [radius=0.1cm];}}
\newrobustcmd*{\mytriangle}[1]{\tikz{\filldraw[draw=#1,fill=#1] (0,0) -- (0.2cm,0) -- (0.1cm,0.2cm);}}
\usepackage{xcolor}
\definecolor{light-gray}{gray}{0.5}
\definecolor{blue}{rgb}{0.0,0.0,1.0}
\definecolor{green}{rgb}{0.0,0.5,0.0}
\definecolor{red}{rgb}{1.0,0.0,0.0}
\definecolor{cyan}{rgb}{0.0,0.75,0.75}
\definecolor{magenta}{rgb}{0.75,0.0,0.75}
\definecolor{yellow}{rgb}{0.75,0.75,0.0}
\definecolor{orange}{rgb}{0.9,0.3,0.0}

\newcommand{\Nu}{\mathrm{Nu}}
\newcommand{\avg}[1]{\langle{#1}\rangle}

\newcommand{\what}{\widehat}

\begin{document}

\preprint{APS/123-QED}

\title{Zonal flow reversals in two-dimensional Rayleigh--B\'enard convection}
\author{P.~Winchester}
\email{winchester@maths.ox.ac.uk}
\author{V.~Dallas}
\email{vassilios.dallas@maths.ox.ac.uk}
\author{P.~D.~Howell}
\email{howell@maths.ox.ac.uk}
\affiliation{Mathematical Institute, University of Oxford, Oxford, OX2 6GG, UK}
\date{\today}

\begin{abstract}
We analyse the nonlinear dynamics of the large scale flow in Rayleigh--B\'enard convection in a two-dimensional, rectangular geometry of aspect ratio $\Gamma$. We impose periodic and free-slip boundary conditions in the streamwise and spanwise directions, respectively. As Rayleigh number Ra increases, a large scale zonal flow dominates the dynamics of a moderate Prandtl number fluid. At high Ra, in the turbulent regime, transitions are seen in the probability density function (PDF) of the largest scale mode. For $\Gamma = 2$, the PDF first transitions from a Gaussian to a trimodal behaviour, signifying the emergence of reversals of the zonal flow where the flow fluctuates between three distinct turbulent states: two states in which the zonal flow travels in opposite directions and one state with no zonal mean flow. Further increase in Ra leads to a transition from a trimodal to a unimodal PDF which demonstrates the disappearance of the zonal flow reversals. On the other hand, for $\Gamma = 1$ the zonal flow reversals are characterised by a bimodal PDF of the largest scale mode, where the flow fluctuates only between two distinct turbulent states with zonal flow travelling in opposite directions.
\end{abstract}

\maketitle
Large scale zonal flow in buoyancy-driven convection is found in the atmosphere of Jupiter \cite{Heimpel05, Kong18, Kaspi18}, in the Earth's oceans \cite{Max05, Richardson06, Nadiga06}, in nuclear fusion devices \cite{diamondetal05, fujisawa08}, in laboratory experiments \cite{Zhang20, Read15, Krishnamurti1981}, and recently in numerical simulations of two-dimensional (2D) Rayleigh--B\'enard convection \cite{Gol14, VDP14, Lohse20}. This large scale flow can undergo abrupt transitions, seemingly randomly, after very long periods of apparent stability \cite{schmeitsdijkstra01,marcus04}.

Such transitions have been observed in a wide range of turbulent flows, including flow past bluff bodies \cite{wygnanskietal86,cadotetal15}, von K\'{a}rm\'{a}n flow \cite{labbeetal96,raveletetal04,torreburguete07}, reversals in a dynamo experiment \cite{berhanuetal07}, Rayleigh--B\'{e}nard convection \cite{stevensetal09,weietal15}, Taylor--Couette flow \cite{huismanetal14}, experiments on 2D turbulence \cite{micheletal16} and Kolmogorov flow \cite{dsf20}. In the turbulent regime, the broken symmetries of the flow can be restored statistically \cite{frisch95}. However, these flows undergo transitions within the turbulent regime, which lead to different flow states as a control parameter increases, and correspond to spontaneous symmetry breaking in a system far from equilibrium,.
These results contradict the idea of a universal state for fully developed turbulence, where, according to Kolmogorov~\cite{K41}, the fluctuations are so vigorous that the system explores the whole high-dimensional phase-space.

In Rayleigh--B\'enard convection, such transitions have also been observed in the form of reversals of the large scale flow in various set-ups \cite{sreenivasanetal02,sugiyamaetal10,nietal15,xieetal18}. In particular, reversals of the large scale circulation in an enclosed rectangular geometry have been observed in experiments and numerical simulations \cite{breuerhansen09,Yanagisawa11,petscheletal11,chandraverma13,podvinsergent15,Verma15}.

In this letter, we report on the reversals of the large scale zonal flow that emerge in the turbulent regime of 2D Rayleigh--B\'enard convection. These transitions occur between long-lived metastable states on a fluctating background, and thus resemble phase transitions in condensed matter physics \cite{Kadanoff20}. Thus, the present work could be of interest to a wider range of fields beyond fluid dynamics. Moreover, the zonal flow reversals are found in the classical Rayleigh--B\'enard convection set-up of a rectangular geometry, with periodic and free-slip boundary conditions in the streamwise and spanwise directions respectively, encouraging further theoretical developments on this idealised set-up.

We adopt the Boussinesq approximation, assuming constant kinematic viscosity $\nu$, and thermal diffusivity $\kappa$. The resulting equations governing 2D Rayleigh--B\'enard convection, written in terms of the stream function $\psi(x, y, t)$ and the perturbation $\theta(x, y, t)$ from the steady state temperature, are
\begin{align}
    \label{psi}
    \psi_t + \nabla^{-2}\{\psi,\nabla^2 \psi \}&=  g \alpha \nabla^{-2}\theta_x + \nu \nabla^2 \psi, \\
    \label{theta}
    \theta_t + \{\psi, \theta \} &= \frac{\Delta T}{\pi d} \psi_x + \kappa \nabla^2 \theta,
\end{align}
where $\{f, g\} = f_x g_y - g_x f_y$ is the standard Poisson bracket. Our spatial domain is bounded vertically by $y \in [0,\pi d]$ and horizontally by $x \in [0,2\pi L]$.
The imposed boundary conditions are periodic in $x$ and free-slip in the $y$-direction, specifically $\psi(x,y,t) = \psi(x+ 2\pi L,y,t)$, $\theta(x,y,t) = \theta(x+ 2\pi L,y,t)$ and $\psi = \psi_{yy} = \theta = 0$ at $y = 0,\,\pi d$.
The three non-dimensional parameters are the aspect ratio $\Gamma = 2 L /  d$, the Prandtl number $\text{Pr} = \nu/\kappa$, and the Rayleigh number $\text{Ra} = \alpha g\Delta T (\pi d)^3 /(\nu \kappa)$, where $\alpha$ is the thermal expansion coefficient, $\Delta T$ is the temperature difference between the top and bottom plates and $g$ is the acceleration due to gravity. In this study, we fix $\text{Pr} = 30$ and consider $\Gamma = 1$ and 2 for Ra ranging from $10^4$ to $10^7$.

\begin{figure*}[!ht]
\centering
\begin{subfigure}{.49\textwidth}
  \centering
  \includegraphics[width=1\linewidth]{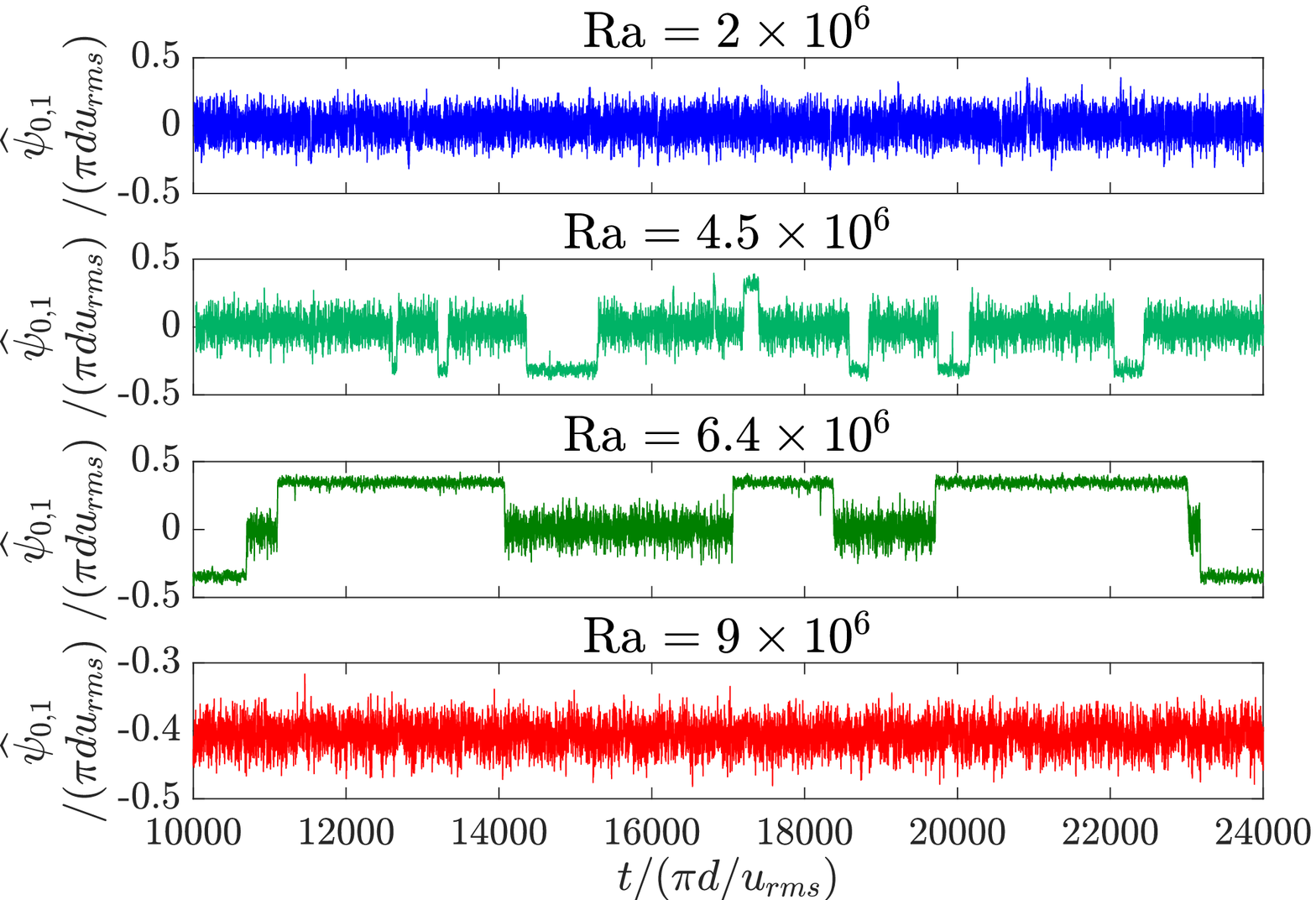}
  \caption{}
  \label{TimeSer}
\end{subfigure}
\begin{subfigure}{.49\textwidth}
  \centering
  \includegraphics[width=1\linewidth]{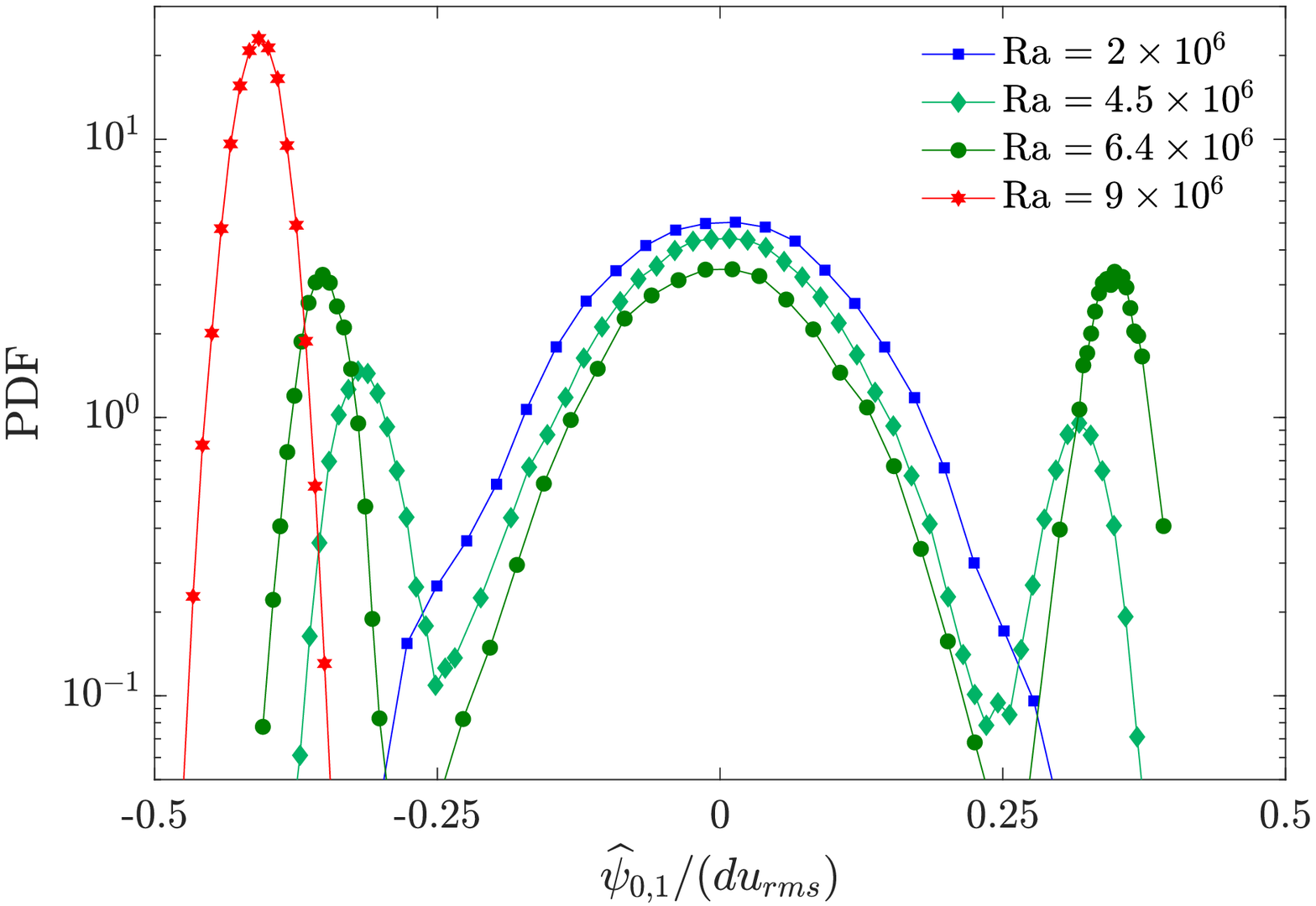}
    \caption{}
  \label{fig:PDF}
\end{subfigure}
\caption{ (Color online) (a) Time series of the normalised large scale mode $\what \psi_{0,1}$ for $\Gamma = 2$ 
and (b) their corresponding PDFs for different values of Ra.}
\label{TimePDF}
\end{figure*}

We perform direct numerical simulations (DNS) by integrating Eqs.~\eqref{psi} and \eqref{theta} using the pseudospectral method \cite{Orszag77}. Based on the numerical code from \cite{dsf20,sdf20b} we decompose the stream function into basis functions with Fourier modes in the $x$-direction and sine modes in the $y$-direction, viz.
\begin{align}
    \psi(x,y,t) &= \sum^{N_x/2}_{k_x = -N_x/2} \sum^{N_y}_{k_y = 1} \widehat{ \psi}_{k_x,k_y}(t) e^{ik_x x/L} \sin{(k_y y/d)},
\end{align}
where $\widehat{ \psi}_{k_x,k_y}$ is the amplitude of the ($k_x, k_y$) mode of $\psi$, and ($N_x, N_y$) denotes the number of aliased modes in the $x$- and $y$-directions. We decompose $\theta$ in the same way. A third-order Runge-Kutta scheme is used for time advancement and the aliasing errors are removed with the two-thirds dealiasing rule \cite{mpicode05b}. For $\text{Ra} < 10^6$ a resolution of $N_x = N_y = 128$ is used. For $\text{Ra} \geq 10^6$, we increase the resolution to $N_x = N_y = 256$. All our runs were integrated to at least $10^4$ eddy turnover times. Integrations for such very long times are necessary to accumulate reliable statistics for the zonal flow transitions. To verify our findings, some runs were repeated at a finer resolution.

We are interested in quantifying the transitions of the large scale flow as the Rayleigh number is increased. Thus, we consider the largest scale mode $\what \psi_{0,1}(t)$, defined as
\begin{align}
    \what \psi_{0,1}(t) = \frac{1}{\pi^2 L d} \int_0^{2\pi L} \int_0^{\pi d} \psi(x,y,t)\sin{( y/d)} \, dy dx.
\end{align}
The emergence of $\what \psi_{0,1}$ spontaneously breaks the centreline symmetry about $y = \pi d/2$ to form a zonal mean profile.

We first focus on simulations of an anisotropic domain with $\Gamma = 2$.
Time series of $\what \psi_{0,1}$ normalised appropriately
(using the depth $\pi d$ and the rms velocity \mbox{$u_{\text{rms}} =  \avg{\lvert\boldsymbol{\nabla}\psi\rvert^2}^{1/2}_{\mathbf{x},t}$} where $\avg{\cdot}_{{\bf x},t}$ denotes a spatio-temporal average) and their corresponding probability density functions (PDFs) are displayed in Fig.~\ref{TimePDF} for different values of Ra.

At $\text{Ra} = 2 \times 10^6$, the time series is turbulent with the amplitude of $\what \psi_{0,1}$ fluctuating randomly around the zero mean. This is an example of non-shearing convection as $\what \psi_{0,1}$ does not break the centreline symmetry in a statistical sense, i.e. $\avg{\what \psi_{0,1}} = 0$, where $\avg{\cdot}$ denotes a time average. The PDF of this time series (blue squares) is close to Gaussian. For non-shearing convection, the resulting flow is characterised by the usual convection rolls.

At $\text{Ra} = 4.5 \times 10^6$, the PDF (green diamonds) has three distinct peaks. This follows the first bifurcation where the system transitions from an approximate Gaussian to a trimodal distribution with two symmetric maxima either side of the peak around $\what \psi_{0,1} = 0$.
This behaviour is related to the emergence of two symmetric states, and the time series is characterised by abrupt and random transitions between these two states (i.e.\ $\avg{\what \psi_{0,1}} > 0$ and $\avg{\what \psi_{0,1}} < 0$) and the non-shearing state (i.e.\ $\avg{\what \psi_{0,1}} \approx 0$).

For $\text{Ra} = 6.4 \times 10^6$ we get random reversals of the large scale flow with a PDF (green circles) which is again trimodal. The system now spends longer intervals in the symmetric shearing states, and the correspondong peaks in the PDF are therefore stronger compared with the peak at $\avg{\what \psi_{0,1}} \approx 0$. As we keep increasing Ra, the reversals become rarer until we get to a transition where no more large scale flow reversals are observed. This is seen in the case with $\text{Ra} = 9 \times 10^6$ where the system was never observed to reverse, remaining stuck in one of the shearing states, and the corresponding PDF (red hexagons) is unimodal with a non-zero mean. The PDF for the largest $\text{Ra} = 9 \times 10^6$ chooses either a positive or negative value of $\what \psi_{0,1}$ depending on the initial condition. The unimodal distribution indicates that ensemble averaging is not equivalent to time averaging for this system.

\begin{figure}[t]
    \includegraphics[width=1\linewidth]{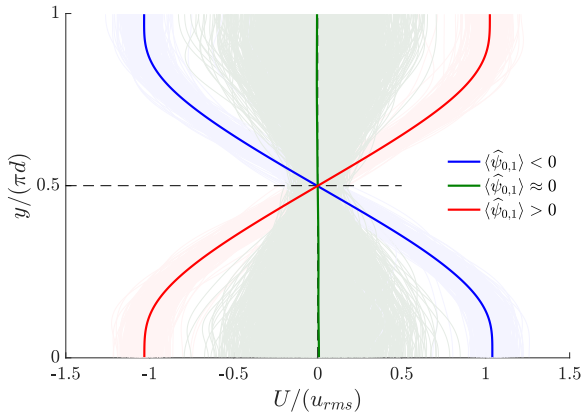} 
    \caption{\label{Revs} (Color online) Time averaged zonal mean flow profiles when $\langle \what \psi_{0,1} \rangle < 0$ (blue), $\langle \what \psi_{0,1} \rangle \approx 0$ (green) and $\langle \what \psi_{0,1}\rangle > 0$ (red) for $\text{Ra} = 6.4 \times 10^6$ and $\Gamma = 2$. The light coloured curves represent instantaneous zonal mean flow profiles and the thicker curves are the averages of these.}
\end{figure}

To understand the flow structure of the different states, in Fig.~\ref{Revs} we plot the zonal mean flow profile
\begin{align}
 U(y,t) = -\frac{1}{2\pi L}\int_{0}^{2\pi L} \psi_y(x,y,t) \, dx
\end{align}
normalised with $u_{rms}$, for the flow with $\text{Ra} = 6.4 \times 10^6$.
The light coloured curves indicate instantaneous realisations of the zonal mean flow profile at different times. These times correspond to $\avg{\what \psi_{0,1}} < 0$ for the light-blue curves, to $\avg{\what \psi_{0,1}} > 0$ for the light-red curves and to $\avg{\what \psi_{0,1}} \approx 0$ for the light-green curves.
The thicker blue, red and green curves are the time averages of the corresponding light-coloured curves. The shear developed in the two shearing states is anti-symmetric with respect to the centreline
$y/(\pi d) = 1/2$. When $\langle \what \psi_{0,1}\rangle > 0$, we observe strong eastward and westward moving flow in the upper half ($1/2 < y/\pi d < 1$) and lower half ($0 < y/\pi d < 1/2$) of the domain respectively. The opposite is true when $\langle \what \psi_{0,1}\rangle < 0$, while there is no time-averaged zonal mean flow when $\langle \what \psi_{0,1} \rangle \approx 0$ even though some instantaneous profiles can be considered having fairly strong shear due to the fluctuations of $\what \psi_{0,1}$ around zero.The transition between the two shearing states captures the reversals of the large scale zonal flow, which occur on a time scale much longer than the eddy turnover time. 

\begin{figure}[t!]
    \includegraphics[width=1\linewidth]{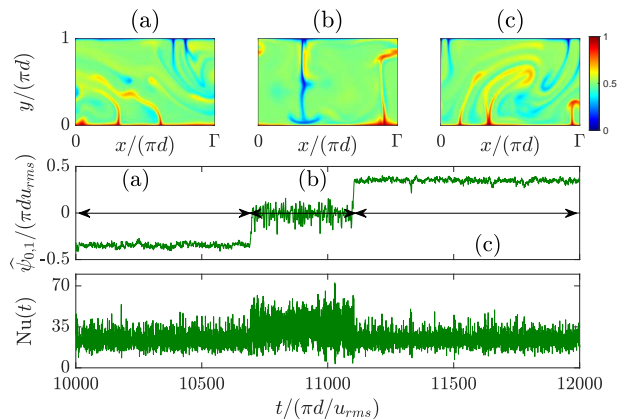}
    \caption{\label{Nu} (Color online) The top row of figures shows instantaneous realisations of the temperature field $T = \Delta T(1-y/\pi d) + \theta$ for $\text{Ra} = 6.4 \times 10^6$ and $\Gamma = 2$. The middle and bottom rows have associated time series for $\what \psi_{0,1}$ and ${\Nu}(t)$ respectively, with times (a), (b) and (c) annotated.}
\end{figure}

To quantify the instantaneous heat transport we consider the time-dependent Nusselt number,
\begin{align}
    \Nu(t) = 1 + \frac{\pi d}{\kappa \Delta T} \avg{\theta \psi_x}_{\bf x},
\end{align}
where $\avg{.}_{\bf x}$ denotes a spatial average. Within the regime of zonal flow reversals, we observe significant reduction in the heat transport whilst the system is in a shearing state. Fig.~\ref{Nu} shows instantaneous realisations of the temperature field $T = \Delta T(1-y/\pi d) + \theta$ along with the time series of the normalised $\what \psi_{0,1}$ and $\Nu$ at $Ra = 6.4 \times 10^6$.
In shearing states such as (a) or (c), the shear prevents thermal plumes from traversing the domain, decreasing the convective heat transport. In non-shearing states such as (b), the shear is not strong enough to suppress thermal plumes and convection rolls are sustained promoting convective heat transport. This observation is supported further by the time average values that the Nusselt number takes in Fig.~\ref{Nu} depending on when the system is in a shearing or non-shearing state, with $\avg{\Nu} = 24.3$ or 34, respectively.

\begin{figure}[!ht]
 \includegraphics[width=1\linewidth]{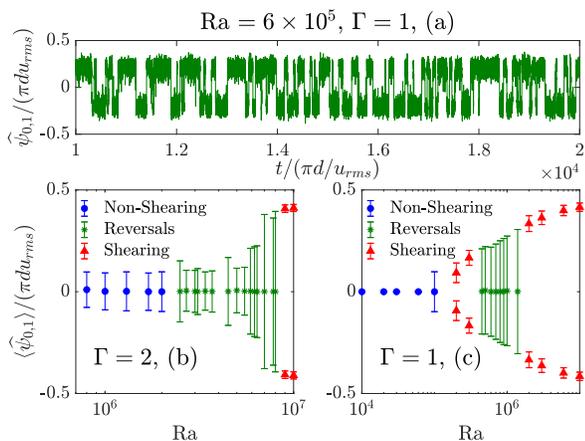}
 \caption{\label{Bif} (Color online)
 (a) Time series of $\what \psi_{0,1}$ for $\text{Ra} = 6 \times 10^5$ and $\Gamma = 1$.
 (b) \& (c) Bifurcation diagrams for $\Gamma = 2$ and $\Gamma = 1$, respectively. Error bars show one standard deviation in the time series of $\what \psi_{0,1}$.
 The non-shearing, shearing and reversing regimes are highlighted by (\mycircle{blue}), (\mytriangle{red}) and (\textcolor{green}{$*$}), respectively.}
\end{figure}

Now, we explore the effect of the aspect ratio on the transitions of the large scale mode by considering an isotropic domain with $\Gamma = 1$. At $\text{Ra} = 6 \times 10^5$ the time series of $\what \psi_{0,1}/(\pi d u_{rms})$ is characterised by abrupt and random transitions between two symmetric shearing states (see Fig.~\ref{Bif}(a)).
In contrast with the case where $\Gamma=2$, we no longer observe a transition to a non-shearing state within the  regime that zonal flow reversals occur. In addition, the bifurcation diagrams of $\avg{\what \psi_{0,1}}$ for $\Gamma = 2$ (Fig.~\ref{Bif}(b)) and $\Gamma = 1$ (Fig.~\ref{Bif}(c)) demonstrate that the nature of bifurcations with respect to the Rayleigh number depends on the aspect ratio.

For $\Gamma=2$, we observe two bifurcations in the system as $\mathrm{Ra}$ increases. As we have already seen, the first bifurcation designates the onset of random zonal flow reversals between two symmetric shearing states and a non-shearing state, and is characterised by the transition from a Gaussian to a trimodal distribution for the time series of $\what \psi_{0,1}$ (see Fig.~\ref{fig:PDF}). The second bifurcation is characterised by the transition of a trimodal to a one-sided unimodal distribution, which designates the disappearance of zonal flow reversals in the system. On the other hand, for $\Gamma = 1$ we observe three bifurcations (see Fig.~\ref{Bif}(c)). In this case, the first bifurcation occurs from a non-shearing state to a persistent shearing state and this is characterised by a transition from a Gaussian to an one-sided unimodal distribution of $\what \psi_{0,1}$. The second bifurcation for $\Gamma = 1$ is the one that designates the onset of random zonal flow reversals. However, note here that the transition is from an one-sided unimodal distribution to a bimodal distribution of $\what \psi_{0,1}$ as it can be inferred from Figs. \ref{Bif}(a) and \ref{Bif}(c). Finally, the third bifurcation for $\Gamma = 1$ is the one that designates the disappearance of zonal flow reversals with a transition from a bimodal to an one-sided unimodal distribution. Note that when $\Gamma = 1$ the regime of zonal flow reversals occurs at values of Ra which are an order of magnitude smaller than when $\Gamma = 2$. Moreover, as the aspect ratio decreases, shearing states emerge at lower Rayleigh numbers (see Fig.~\ref{Bif}(c)), in agreement with recent results \cite{Lohse20}.

In summary, for a moderate Prandtl number fluid in 2D Rayleigh--B\'enard convection we observe the emergence of a large scale zonal flow, whose dynamics are dominated by the largest scale mode $\what \psi_{0,1}$. As the Rayleigh number increases, we find large scale flow transitions between long-lived metastable states within the turbulent regime of the system. These transitions are seen in the PDF of the time series of $\what \psi_{0,1}$ mode. For aspect ratio $\Gamma = 2$, the PDF transitions first from a Gaussian to a trimodal distribution signifying the onset of reversals between two symmetric shearing states and a non-shearing state. The zonal flow reversals suppress the convective heat transfer as thermal plumes are not able to traverse the layer. Then, as Ra increases further, a second transition occurs from a trimodal to an one-sided unimodal distribution, where reversals cease to exist for the whole duration of the simulation. For $\Gamma = 1$ similar flow transitions are observed but the reversals in this case happen between two symmetric shearing states giving a bimodal PDF for the large scale mode. Quantitive understanding of such transitions over fluctuating background remains a challenge for the future. Low-order models derived from the governing equations is one way to theoretically understand further these dynamics.

\acknowledgements{We would like to thank K. Seshasayanan for his comments on an initial version of this manuscript.}

\bibliography{references}

\end{document}